\documentclass[twoside,slac_one]{revtex4}
\usepackage{graphicx}
\usepackage{fancyhdr}
\usepackage{amsmath} 
\usepackage{bm}
\usepackage{amsxtra}
\usepackage{amssymb}
\usepackage{amsthm}
\usepackage{latexsym}
\usepackage{lscape}

\begin{document}

\title{The Charge Radius of the Proton}

%

\author{G. Paz}
\affiliation{Department of Physics and Astronomy, Wayne State University, Detroit, MI, USA}

\begin{abstract}
Recently, the charge radius of the proton was extracted for the first time from muonic hydrogen. The value obtained is five standard deviations  away from similar measurements of regular hydrogen. This talk discusses work done in collaboration with Richard J. Hill, to address this discrepancy. First, we have studied the extraction of the charge radius of the proton from electron-proton scattering data in a model-independent way. We have shown that previous extractions, spanning a period of over 40 years, have underestimated their errors. Second, we have looked at a model-independent analysis of proton structure effects for hydrogen-like bound states, using the tool of an effective field theory, namely NRQED. We have identified hidden model-dependent assumptions in the theoretical calculation behind the muonic hydrogen result.
\end{abstract}

\maketitle

\thispagestyle{fancy}


\section{Introduction}
How big is the proton? To answer this question we first have to define what do we mean by the proton's size. The most common definition is of the charge radius of the proton. The matrix element of the electromagnetic current between nucleon states 
gives rise to two form factors,  
\begin{equation}
\langle N(p_f)|\sum_q\,e_q\,\bar q\gamma^\mu q|N (p_i)\rangle=
\bar{ u}(p_f)
\left[\gamma_\mu {F_1(q^2)}+\frac{i\sigma_{\mu\nu}}{2m}{ F_2(q^2)}q_\nu\right]\hspace{-0.2em}u(p_i),
\end{equation}
where $q=p_f-p_i$ and $m$ is the nucleon mass. $F_1$ and $F_2$ are often referred to as the Dirac and Pauli form factors, respectively.  The Sachs electric and magnetic form factors are defined as $G_E(q^2) = F_1(q^2) + {q^2} F_2(q^2)/4m^2 $,  $G_M(q^2) = F_1(q^2) + F_2(q^2)$. The values of $G_E$ and $G_M$ at $q^2=0$ give the electric charge and magnetic moment of the nucleon, respectively. For example, $G^p_E(0) = 1$ and $G^p_M(0) = \mu_p \approx 2.793$. The electric charge radius of the proton is defined by the slope of $G^p_E$
\begin{equation}\label{rdef}
\left(r_E^p\right)^2 =6\frac{dG^p_E}{dq^2}\Bigg|_{q^2=0}. 
\end{equation}
The motivation for this definition is that in the Breit, or brick-wall, frame, $r_E^p$ is the root-mean-square radius of the charge distribution. Notice, though, that the definition (\ref{rdef}) is independent of any specific frame.

 The most direct way to extract the charge radius of the proton is by measuring $G^p_E$ from  scattering experiments, usually electron-proton scattering. An indirect way is to use spectroscopy of a bound proton-lepton system. Intuitively, the fact that the proton is not a point particle but has a structure, modifies the potential seen by the leptons. For a point particle the amplitude for $p+\ell \to p+\ell$ is proportional to $1/q^2$, leading to the well known $-Z\alpha/r$ potential. Including  $q^2$ corrections from proton structure lead to an amplitude proportional to a constant.  In position space it gives a delta function type of a potential, proportional to the charge radius squared, $4\pi Z \alpha\delta^3(r)(r_E^p)^2/6$. The corresponding correction to the energy level is 
$\Delta E_{r^p_E}={2(Z\alpha)^4}{ m_r^3}(r_E^p)^2 \delta_{\ell\,0}/3n^3\,,$
where  $m_l$ is the lepton mass ($l=e,\mu$), $m_r=m_\ell m_p/(m_\ell+m_p)$, and $m_p$ is the proton mass. For regular and muonic hydrogen $m_r \approx m_\ell$. Since $m_\mu\gg m_e$,  proton structure effects for muonic hydrogen are greatly enhanced compared to regular hydrogen. 

For a long time it was anticipated that a measurement of the Lamb shift in muonic hydrogen would reduce the error by an order of magnitude compared to measurements from electron-proton scattering and regular hydrogen spectroscopy \cite{Pachucki:1996zza}. Last year, the CREMA collaboration at the Paul Scherrer Institute in Switzerland have reported the first measurement of the charge radius of the proton from the Lamb shift in muonic hydrogen \cite{Pohl:2010zz}. The value obtained was $r_E^p = 0.84184(67)$ fm. As expected, the error is an order of magnitude smaller than the value of $r_E^p =0.8768(69)$~fm, extracted mainly from (electronic) hydrogen spectroscopy  by CODATA (Committee on Data for Science and Technology) \cite{Mohr:2008fa}. But the central value is five standard deviation away! It is quite rare in high energy physics to have a five sigma deviation, which motivated several studies of new physics effects \cite{Barger:2010aj, TuckerSmith:2010ra, Batell:2011qq}. How should we interpret this result?

\section{Model independent extraction of the proton charge radius from electron scattering} 

Since we are having two conflicting measurements, it is useful to look at a third one, the measurement from electron-proton scattering data. What is the value extracted from that measurement? The first place to look is the particle data book. The 2010 edition  \cite{Nakamura:2010} lists 12 extractions from electron-proton scattering data that span the period of 1963-2005 and the range of $0.8-0.9$ fm, most with quoted uncertainties of 0.01-0.02 fm. These are based on different data sets and different extraction methods. It is hard to draw any conclusion from this list. Indeed, the Particle Data Group (PDG) itself does not use any of these extractions ``...for averages, fits, limits, etc." 

Part of the problem is the use of different data sets. This problem can be overcome by a world effort in combining data, perhaps in a way similar to the Heavy Flavor Averaging Group \cite{HFAG}. A more subtle issue is the question of how to extract the charge radius of the proton from the measured values of $G_E^p(q^2)$. This seems like a simple problem. We have a data set, and we just need to find the slope at $q^2=0$. The problem is that we do not know the functional form of $G_E(q^2)$, so the extrapolation to $q^2=0$ is not as simple. Since we do not know the functional form, one would like to extract the charge radius in a model-independent way. This rules out some extractions based on a specific form of the form factor, typically a sum of  poles and continuum states. What we would like to have is a general parametrization, a series expansion for example,  that gives a consistent result for the charge radius as we fit more parameters to the data. 

\begin{figure*}[ht]
\centering
\includegraphics[height=0.2\textwidth]{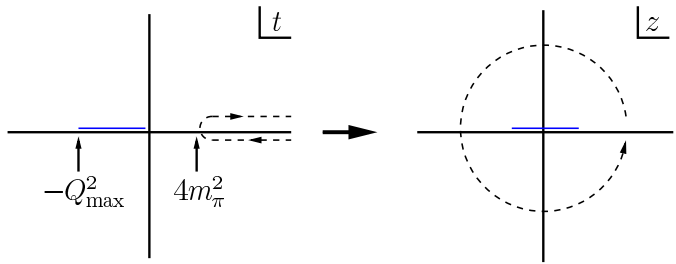}
\caption{Conformal mapping of the cut plane to the unit circle.}
 \label{figure1}
\end{figure*}

There are several series expansion that are used. Perhaps the simplest one is to expand $G_E(q^2)$ in a taylor series around $q^2=0$. Many of the sources mentioned by the PDG have used such an expansion, but typically with a small (2-3) number of parameters. Another type of a series expansion is the continued fraction expansion suggested in \cite{Sick:2003gm}. In \cite{Hill:2010yb}, we have advocated the use the so called ``$z$-expansion", which is a standard tool in analyzing meson transition form factors, (for references see  \cite{Hill:2010yb}). The $z$-expansion utilizes the known analytic properties of $G_E^p(t)$, ($t\equiv q^2$): it is analytic in the complex $t$-plane outside a cut of $t\in[4m_\pi^2,\infty]$. The electron-proton  scattering data lies on the real $t<0$ axis. It is therefore common to define $Q^2=-q^2=-t$. One can map the domain of analyticity onto the unit circle by the conformal transformation 
\begin{equation}
z(t,t_{\rm cut},t_0) = {\sqrt{t_{\rm cut} - t} - \sqrt{t_{\rm cut} - t_0} \over \sqrt{t_{\rm cut} - t} + \sqrt{t_{\rm cut} - t_0}  } \,, 
\end{equation}
where $t_{\rm cut}=4m_\pi^2$ and $t_0$ is a free parameter representing the point mapping onto $z=0$, see figure \ref{figure1} . It is convenient to choose $t_0=0$, but physical observables are independent of the choice of $t_0$. The maximum size of $z$ depends on the distance between $t_{\rm cut}$ and $t=0$. This implies that if we can increase $t_{\rm cut}$, the maximum size of $z$ decreases and the convergence of the series is improved. Since $G_E^p$ is analytic in the unit disc of the $z$ plane, it can be expanded as a taylor series in $z$: $G_E^p(q^2) = 
{\sum_{k=0}^\infty a_k} \, z(q^2)^k$. 

Does it matter which expansion we use? This question was studied in \cite{Hill:2010yb}. Using the low $q^2$ data tabulated in \cite{Rosenfelder:1999cd}, we have compared the extraction of the charge radius of the proton using 4 different expansions: Taylor, continued fraction, $z$ expansion, and  $z$ expansion with a bound on the coefficients, $|a_k|\le10$. While fits of the different expansion with 2 parameters agree well, as we increase the numbers of parameters, the errors for the first 3 fits grow without bound. Only the  constrained $z$ expansion is stable and therefore model-independent. Another interesting feature is that the error on the extracted charge radius for a fit with 2 parameters is actually an underestimate. The final answer has an error that can be larger by a factor of 2. The conclusion is that many previous extractions have underestimated their errors. 

Since the $z$ expansion is  related to the analytic structure of the form factor, information  about  the imaginary part of $G_E^p(t+i0)$ directly translates into information about the expansion coefficients.  The analysis of \cite{Hill:2010yb} had shown that a bound of $|a_k|\le10$ is very conservative. Using the electron-proton scattering data tabulated in \cite{Arrington:2007ux}, we have found that $r_E^p=0.870 \pm 0.023 \pm 0.012$ fm, where the first error 
is obtained using  $|a_k|\le 5$, and the additional error 
for a fit assuming $|a_k|\le 10$.  The errors might seem large compared to some of the values listed in \cite{Nakamura:2010}, but as we have shown in  \cite{Hill:2010yb}, this can be attributed to the model-dependence of the previous extractions.

The error can be reduced by including neutron and  $\pi\pi$ data. The inclusion of neutron data allows for a separate fit of the isoscalar and isovector components of $G_E^p$. For the isoscalar component, the cut starts at $t=9m_\pi^2$. By including the $\pi\pi$ data, namely $\pi\pi$ production and  $\pi\pi \to N\bar{N}$, we can effectively raise the isovector threshold to $t=16m_\pi^2$. As explained before, these typically lead to a better convergence and a smaller error. The best value obtained in 
 \cite{Hill:2010yb} is $r_E^p = 0.871 \pm 0.009 \pm 0.002 \pm 0.002$ fm, where  the first and the second error 
are as above, and the final error is from the $\pi\pi$ continuum contribution, see \cite{Hill:2010yb}  for details. 

The problem of model-dependent extraction of such basic non-perturbative parameters is not unique to the vector form factor. The matrix element of the axial current between nucleon states gives rise to the axial form factor. Analogously to (\ref{rdef}), its slope at $q^2=0$ defines the nucleon axial radius $r_A$, or alternatively, the axial mass defined as $m_A=\sqrt{12}r_A^{-1}$. Extractions of the axial mass usually use a dipole model for the axial form factor, $ F_A=F_A(0)\,[1-q^2/(m_A^{\rm dipole})^2]^{-2}$. Recently,  MiniBooNE has extracted the axial mass from neutrino scattering data. The value obtained, $m_A^{\rm dipole}= 1.35\pm 0.17$  GeV \cite{AguilarArevalo:2010zc}, is in conflict with the value obtained from pion electro-production data, {$m_A^{\rm dipole}=1.07\pm 0.02$  GeV \cite{Bernard:2001rs}. Notice the superscript  ``dipole", emphasizing that fact that a dipole model was used. Using a model-independent approach \cite{Bhattacharya:2011ah}, we have obtained the value $m_A= 0.85^{+0.22}_{-0.07}\pm 0.09$  GeV. From an illustrative dataset for 
pion electro-production, we have extracted the value, $m_A = 0.92^{+0.12}_{-0.13}\pm 0.08 \,{\rm GeV}$. As a check, we have also extracted $m_A^{\rm dipole}$ from the same data sets by using the dipole model. We obtain  
$m_A^{\rm dipole}=1.29\pm 0.05 \,{\rm GeV}$ (neutrino scattering) and 
$m_A^{\rm dipole}=1.00\pm 0.02 \,{\rm GeV}$ (pion electro-production). As far as the axial mass is concerned, the discrepancy can be attributed to the use of dipole model. 

Returning to the issue of the charge radius of the proton, the value obtained from a model-independent extraction from scattering data is 
$r_E^p = 0.871 \pm 0.009 \pm 0.002 \pm 0.002$ fm. It is more consistent with the CODATA value,  $r_E^p =0.8768(69)$ fm, than the CREMA value, $r_E^p = 0.84184(67)$ fm. Let us look more carefully at the muonic hydrogen extraction. 

\section{Model independent analysis of proton structure for hydrogen-like bound states}
The CREMA collaboration has not measured the charge radius of the proton directly. They have measured the Lamb shift of 
$\Delta E\equiv E[2P_{j=3/2}^{(F=2)}]-E[2S_{j=1/2}^{(F=1)}]=206.2949 \pm 0.0032\,\,{\rm meV}$, between the $2S_{j=1/2}^{(F=1)}$ and $2P_{j=3/2}^{(F=2)}$ energy levels. To extract the charge radius it should be compared to the theoretical expression \cite{Pachucki:1999zza, Borie:2004fv}
\begin{equation}\label{cubic}
\Delta E =209.9779(49) - 5.2262 (r_E^p)^2+0.0347(r_E^p)^3\,\,{\rm meV}\,,
\end{equation}
to obtain $r_E^p=0.84184(67)$ fm.  The constant term arises from many contributions, mostly QED effects related to the muon. We have explained in the introduction how  the quadratic term arises. Where does the cubic term come from? 

\begin{figure*}[ht]
\centering
\includegraphics[height=0.2\textwidth]{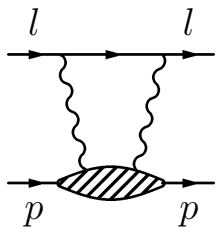}
\caption{Two photon exchange}
 \label{figure2}
\end{figure*}

In the calculation that leads to (\ref{cubic}), the cubic term arises from a two photon exchange between the lepton and the proton, see figure 2. To obtain the cubic term, it is separated to a ``proton" and ``non-proton" contributions.  To calculate  the proton part,  one uses ``Feynman rules" in which onshell form factors are inserted into the vertices of Feynman diagrams for a relativistic pointlike particle. The proton part of the amplitude is now a function of $G_E$ and $G_M$. Assuming a dipole model for the form factors $G_i=G_i(0)\,[1-q^2/\Lambda^2]^{-2}$, the amplitude can be expressed as  function of one parameter, $\Lambda$. Since one parameter model for the form factors is assumed,  this parameter can be related to $r_E^p$. ``Expanding" in powers of $r_E^p$, an approximated cubic dependence is obtained.  Not expanding in powers of $r_E^p$, and using $\Lambda^2=0.71$ GeV, it leads to $\Delta E=0.018$ meV \cite{Pachucki:1996zza}. To explain the discrepancy it should be  $0.258(90)$ meV, if using  $r_E^p$ from scattering data \cite{Hill:2010yb}, or $0.311(63)$ meV, if using  $r_E^p$ from spectroscopy \cite{Mohr:2008fa}.

This procedure raises several questions. First, can we separate the two photon amplitude into a ``proton" and ``non-proton" contributions?  Second, why is the insertion of the form factors in vertices valid? Third, even if the insertion were valid, the dependence on the cubic power of the charge radius is only valid for a one parameter model of the form factors and in the large proton mass limit \cite{Pachucki:1999zza} . Since we are discussing precision measurements, a more solid theoretical treatment is clearly needed. 

To study proton structure effects in hydrogen-like bound states in a model-independent way, one can use an effective field theory, namely Non Relativistic QED (NRQED) \cite{Hill:2011wy}.  The NRQED Lagrangian up to order $1/m_p^3$ is \cite{Caswell:1985ui,Kinoshita:1995mt,Manohar:1997qy}
\begin{eqnarray}\label{NRQED}
{\cal L}_p&=& \psi_p^\dagger
  \Bigg\{  i D_t  + {\bm{D}^2 \over 2 m_p}  + {\bm{D}^4 \over 8 m_p^3} +
  c_F e{ \bm{\sigma}\cdot \bm{B} \over 2m_p}   
+ c_D e{ [\bm{\partial}\cdot \bm{E} ]  \over 8 m_p^2}  + i c_S e{ \bm{\sigma}
    \cdot ( \bm{D} \times \bm{E} - \bm{E}\times \bm{D} ) \over 8 m_p^2} 
  \nonumber \\ 
&& + c_{W1}e {  \{ \bm{D}^2 ,  \bm{\sigma}\cdot \bm{B} \}  \over 8 m_p^3}-c_{W2}e {  D^i \bm{\sigma}\cdot
    \bm{B} D^i \over 4 m_p^3 }  + c_{p^\prime p} e { \bm{\sigma} \cdot
    \bm{D} \bm{B}\cdot \bm{D} + \bm{D}\cdot\bm{B} \bm{\sigma}\cdot \bm{D}
    \over  8 m_p^3} 
    + i c_M e { \{ \bm{D}^i ,  [\bm{\partial} \times
    \bm{B}]^i \}
\over 8 m_p^3}\nonumber \\ 
&& +
  c_{A1} e^2{ \bm{B}^2 - \bm{E}^2 \over 8 m_p^3}  - c_{A2} e^2{ \bm{E}^2
    \over 16 m_p^3 } +...
 \Bigg\} \psi_p  \,,
 \end{eqnarray}
We also need the contact interaction
 \begin{equation}\label{contact}
  {\cal L}_{\rm contact} = d_1
{\psi_p^\dagger \bm{\sigma} \psi_p \cdot  \psi_l^\dagger \bm{\sigma} \psi_l \over m_l m_p} 
+ d_2 {\psi_p^\dagger \psi_p
\psi_l^\dagger  \psi_l \over m_l m_p}\,.
\end{equation}

The knowledge of the Wilson coefficients allows us  to determine the proton structure corrections to the energy levels. Thus
\begin{equation}
\delta E(n,\ell) =  \delta_{\ell 0} {m_r^3 (Z\alpha)^3 \over \pi n^3} \left(\frac{Z\alpha\pi}{2m_p^2}c_D^{\rm proton}-{d_2 \over m_l m_p}\right)\,.
\end{equation}
The Wilson coefficients are determined as follows. Coefficients of operators that contain coupling to one photon field, namely the $c_i$'s in (\ref{NRQED}) apart from  $c_{A1}$ and $c_{A2}$, are determined by the derivatives of the form factors, $F_1$ and $F_2$, at $q^2=0$. Coefficients of operators that contain coupling only  to two photons, namely  $c_{A1}$ and $c_{A2}$,  are determined by the forward and backward Compton scattering.  Coefficients of the contact terms, namely  $d_{1}$ and $d_{2}$, are determined by using the zero-momentum limit for the $p+\ell \to p+\ell$ amplitude. The tree level, ${\cal O } (\alpha)$, amplitude is reproduced by the effective field theory, and $d_1$ and $d_2$ receive a non-zero contribution starting at ${\cal O } (\alpha^2)$. Here we consider only the spin-independent coefficient, $d_2$.  The relevant proton matrix element is the forward Compton amplitude
\begin{equation}\label{Wi}
\frac12 \sum_s i \int d^4x\, e^{iq\cdot x} \langle \bm{k},s| T\{ J_{\rm e.m.}^\mu(x) J_{\rm e.m.}^\nu(0) \} | \bm{k},s \rangle 
=
\left( - g^{\mu\nu} + \frac{q^\mu q^\nu}{ q^2} \right) W_1
+ 
\left( k^\mu - \frac{k\cdot q \,q^\mu}{q^2} \right) 
\left( k^\nu - \frac{k\cdot q \, q^\nu }{q^2} \right) W_2\,,
\end{equation}  
where the $W_i$'s are functions of $\nu = 2k\cdot q$ and $Q^2=-q^2$. The matching condition is 
\begin{eqnarray}
&&{4\pi m_r \over \lambda^3} \!-\! {\pi m_r \over 2 m_l m_p\lambda}  \!-\! { 2\pi m_r \over m_p^2\lambda}\left[ F_2(0) 
\!+\! 4m_p^2 F_1^\prime(0) \right]
- {2\over m_l m_p} \bigg[ 
\frac23 + {1\over m_p^2 - m_l^2}\left( m_l^2 \log{m_p\over \lambda} - m_p^2 \log{m_l\over\lambda} \right)
\bigg] 
+{ {\delta d_2 (Z\alpha)^{-2}\over m_l m_p} } \nonumber\\
&&= -
{m_l\over m_p} \int_{-1}^1 dx   \sqrt{1-x^2} 
\int_0^\infty dQ \, {Q^3 \left[ (1+2x^2){W_1}( 2im_pQ x ,Q^2) - (1-x^2) m_p^2 { W_2}( 2im_pQ x, Q^2 ) \right] 
\over (Q^2 + \lambda^2)^2(Q^2 + 4 m_l^2 x^2)}\,,
\end{eqnarray}
where $\lambda$ is the photon mass and $\delta d_2$ denotes the contribution to $d_2$ in addition to the point particle value. To determine $\delta d_2$, we need to know $W_1$ and $W_2$. How can we determine them? 

The imaginary part of $W_1$ and $W_2$ can be related to measured quantities. Inserting a complete set of states to (\ref{Wi}), the imaginary part of the $W_i$'s can be divided to proton and non-proton contribution. The proton contribution can be expressed in terms of  form factors, while the non-proton contribution can be expressed in terms of inelastic structure functions. To reconstruct the $W_i$'s from their imaginary parts, we need to use dispersion relations:
\begin{eqnarray}\label{dispersion}
W_1(\nu,Q^2) &=& W_1(0,Q^2)+{\nu^2\over \pi} \int_{\nu_{\rm cut}(Q^2)^2}^\infty { d\nu^{\prime 2}} 
{ {\rm Im}W_1(\nu^\prime, Q^2) \over \nu^{\prime 2} (\nu^{\prime 2} - \nu^2) } 
\nonumber\\\nonumber\\
W_2(\nu,Q^2) &=&{1\over \pi} \int_{\nu_{\rm cut}(Q^2)^2}^\infty { d\nu^{\prime 2} }
{ {\rm Im}W_2(\nu^\prime, Q^2) \over \nu^{\prime 2} - \nu^2}. 
\end{eqnarray} 
For convergence, $W_1$ requires a subtraction. As a result, one needs to know also $W_1(0,Q^2)$. Currently $W_1(0,Q^2)$ cannot be calculated from first principles or extracted from experiment. This fact introduces an uncertainty that was largely ignored in the literature. 

Can we say anything model-independently about $W_1(0,Q^2)$? We can calculate it  in the large and small limit of $Q^2$. For small $Q^2$ the photon ``sees" the proton almost like an elementary particle. Using NRQED to calculate $W_1(0,Q^2)$ up to ${\cal O}(Q^2)$ (including) \cite{Hill:2011wy}
\begin{equation}\label{low}
W_1(0,Q^2)=2(c_F^2-1)+2\frac{Q^2}{4m_p^2}\left(c_{A_1}+c_F^2-2c_Fc_{W1}+2c_M\right)\,,
\end{equation}
where we have described above how the $c_i$'s are determined. For large $Q^2$ the photon ``sees" the quarks inside the proton.  Using the Operator Product Expansion to calculate $W_1(0,Q^2)$, we find that  $W_1(0,Q^2)\sim1/Q^2$ for large $Q^2$ \cite{Hill:2011wy}.  The intermediate $Q^2$ region is not constrained by existing measurements. This introduces model-dependence into the theoretical prediction. 

We have described above the procedure leading to (\ref{cubic}). We can now address some of the questions that it raises. First, the separation to proton and non-proton states is only justified for the imaginary part of the $W_i$'s. $W_1(0,Q^2)$ cannot be rigorously divided in such a way. Second,  the ``proton part" of $W_1(0,Q^2)$ is evaluated in the literature by inserting onshell form factors into the vertices of Feynman diagrams for a relativistic pointlike particle. The result, which we refer to as the ``Sticking In Form Factors" (SIFF) model, gives $W_1^{\rm SIFF}(0,Q^2) =2F_2(Q^2)\left[2F_1(Q^2) + F_2(Q^2)\right]$. This result is not  derived from a well defined local field theory. In fact, \emph{no} local Lagrangian can give such Feynman rules. Furthermore, the resulting expression for $W_1(0,Q^2)$ has the wrong large $Q^2$ behavior. The ``non-proton"  part is obtained	by multiplying part of the term proportional to $c_{A1}$ in (\ref{low}) by a function of $Q^2$ \cite{Pachucki:1999zza}. This function is not derived from first principles and again does not have the correct large $Q^2$ behavior. We stress again that unlike the imaginary part of the $W_i$'s, the separation of $W_1 (0,Q^2 )$ into proton and non-proton parts is not well-defined. In conclusion, the standard calculation behind  (\ref{cubic}) is model-dependent. One might ask though, how severe is the model dependence?

Considering (\ref{dispersion}),  it is natural to decompose the two photon contribution to the energy levels as $\Delta E^{{\rm two}-\gamma}=\Delta E^{{\rm continuum }}+\Delta E^{{\rm proton}}+\Delta E^{W_1(0,Q^2)}$. The first two terms in this equation can be extracted from data and are therefore model-independent. The last term is model-dependent.  We look at the $n=2$ level of muonic hydrogen. The continuum contribution was recently determined to be \cite{Carlson:2011zd}, $\Delta E^{{\rm continuum }}_{\mu H}=0.0127(5) \mbox{ meV}$.  To illustrate the magnitude of the proton contribution, let us use a simple dipole model: $G_E(q^2) \approx {G_M(q^2)/G_M(0)} \approx [1 - q^2/\Lambda^2]^{-2}$ with $\Lambda^2= 0.71\,{\rm GeV}^2$.  We find $\Delta E^{{\rm proton}}_{\mu H}=-0.016 \mbox{ meV}$. This result can be refined by using the measured values of the form factors. In the SIFF model, using the dipole form factors, we find $\Delta E^{W_1(0,Q^2), {\rm SIFF}}_{\mu H}=0.034 \mbox{ meV}$. The sum of the last two terms gives the $0.018$ meV mentioned above \cite{Pachucki:1996zza}. In other words, the theoretical prediction for the ``proton contribution" consists of a sum of two terms. The model dependent piece is twice as large as the model-independent piece and has the opposite sign. Needless to say, this result is very alarming. The theoretical prediction used for the extraction of the charge radius of the proton from muonic hydrogen relays heavily on the validity of the SIFF model.  It is not hard to construct models of $W_1(0,Q^2)$ that would that have the correct small-$Q^2$ and large-$Q^2$ behavior, but give a much larger contribution than the SIFF model.  

\section{Conclusions}

In this talk we have discussed the recent discrepancy between the extraction the proton charge radius from muonic ($r_E^p = 0.84184(67)$ fm) and regular ($r_E^p =0.8768(69)$~fm) hydrogen. We have looked at the extraction of the charge radius from electron-proton scattering data. Such an extraction is complicated by the fact that the functional form of the form factor is unknown.  We have established in \cite{Hill:2010yb} that previous extractions suffer from model-dependence and that the error on the charge radius was underestimated. This situation is not unique to the vector form factor. The recent discrepancy in the extraction of the axial mass from the  MiniBooNE data is in conflict with previous pion electro-production data, when a one-parameter model for the axial form-factor is used. When a model independent approach is taken,   \cite{Bhattacharya:2011ah}, there is no discrepancy, as far as the axial mass is concerned. We have used a model independent approach to extract the charge radius of the proton from proton, neutron and $\pi\pi$ data. The value of the charge radius we obtain, $r_E^p = 0.871(11)$ fm, is closer to the regular hydrogen result. This motivates a closer look at the muonic hydrogen determination. 

We have presented the NRQED formalism for isolating and controlling proton structure effects in hydrogen-like bound states. The most severe problem with the theory behind the muonic hydrogen measurement arises from the contact interaction (\ref{contact}). The Wilson coefficient $d_2$ is not determined by proton form factors or inelastic structure functions. To determine it,  the two photon amplitude is needed.

 Let us summarizes the discussion of the two photon amplitude. To determine $d_2$ we need the imaginary part of the $W_i$'s which can be extracted from data, and also $W_1(0,Q^2)$ which currently cannot be extracted from data. Unlike the imaginary part, $W_1(0,Q^2)$ \emph{cannot} be written model independently as a sum of ``proton" and ``non-proton" terms. The theoretical calculation behind the  muonic hydrogen result assumes such a separation and relies on the SIFF model to estimate the ``proton" part. Even worse, the SIFF model is not identified as a model, but presented as a theory.  In \cite{Hill:2011wy} we have calculated model-independent properties of $W_1(0,Q^2)$, namely its low $Q^2$ behavior via NRQED and 
high $Q^2$ behavior via OPE. The intermediate region is poorly constrained. The lack of theoretical control over  $W_1(0,Q^2)$ introduces theoretical uncertainties that were not taken into account in the literature.

While not resolving the discrepancy per se, our studies have exposed hidden model-dependence in the extraction of the charge radius of the proton from scattering data and in the theoretical calculation behind the extraction of the charge radius of the proton from muonic hydrogen. Hopefully, these studies will serve as a basis for a more careful extraction of this basic quantity.


\begin{thebibliography}{99} 

\bibitem{Pachucki:1996zza}
  K.~Pachucki,
  Phys.\ Rev.\  A {\bf 53}, 2092 (1996).

\bibitem{Pohl:2010zz}
  R.~Pohl {\it et al.},
  Nature {\bf 466}, 213 (2010).
  
 \bibitem{Mohr:2008fa}
  P.~J.~Mohr, B.~N.~Taylor and D.~B.~Newell,
  Rev.\ Mod.\ Phys.\  {\bf 80}, 633 (2008)
  [arXiv:0801.0028 [physics.atom-ph]].
  
\bibitem{Barger:2010aj}
  V.~Barger, C.~-W.~Chiang, W.~-Y.~Keung, D.~Marfatia,
  Phys.\ Rev.\ Lett.\  {\bf 106}, 153001 (2011)
  [arXiv:1011.3519 [hep-ph]].
  
 \bibitem{TuckerSmith:2010ra}
  D.~Tucker-Smith, I.~Yavin,
  Phys.\ Rev.\  D {\bf 83}, 101702 (2011)
  [arXiv:1011.4922 [hep-ph]].   
  
\bibitem{Batell:2011qq}
  B.~Batell, D.~McKeen, M.~Pospelov,
  Phys.\ Rev.\ Lett.\  {\bf 107}, 011803 (2011)
  [arXiv:1103.0721 [hep-ph]].  
  
  \bibitem{Nakamura:2010}
  K.~Nakamura {\it et al.}  [Particle Data Group],
  J.\ Phys.\ G {\bf 37}, 075021 (2010).
  
  \bibitem{HFAG}
  http://www.slac.stanford.edu/xorg/hfag/
  
  \bibitem{Sick:2003gm}
  I.~Sick,
  Phys.\ Lett.\  B {\bf 576}, 62 (2003)
  [arXiv:nucl-ex/0310008].
  
  \bibitem{Hill:2010yb}
  R.~J.~Hill, G.~Paz,
  Phys.\ Rev.\  D {\bf 82}, 113005 (2010)
  [arXiv:1008.4619 [hep-ph]].
  
\bibitem{Rosenfelder:1999cd}
  R.~Rosenfelder,
  Phys.\ Lett.\  B {\bf 479}, 381 (2000)
  [arXiv:nucl-th/9912031].
  
\bibitem{Arrington:2007ux}
  J.~Arrington, W.~Melnitchouk and J.~A.~Tjon,
  Phys.\ Rev.\  C {\bf 76}, 035205 (2007)
  [arXiv:0707.1861 [nucl-ex]].

\bibitem{AguilarArevalo:2010zc}
  A.~A.~Aguilar-Arevalo {\it et al.}  [MiniBooNE Collaboration],
  arXiv:1002.2680 [hep-ex].
  
  \bibitem{Bernard:2001rs}
  V.~Bernard, L.~Elouadrhiri and U.~G.~Meissner,
  J.\ Phys.\ G {\bf 28}, R1 (2002)
  [arXiv:hep-ph/0107088].
  
 \bibitem{Bhattacharya:2011ah}
  B.~Bhattacharya, R.~J.~Hill, G.~Paz,
  [arXiv:1108.0423 [hep-ph]], to appear in Phys.\ Rev.\  D.
  
\bibitem{Pachucki:1999zza}
  K.~Pachucki,
  Phys.\ Rev.\  A {\bf 60}, 3593 (1999).
  
\bibitem{Borie:2004fv}
  E.~Borie,
  Phys.\ Rev.\  A {\bf 71}, 032508 (2005)
  [arXiv:physics/0410051].
  

 \bibitem{Hill:2011wy}
  R.~J.~Hill, G.~Paz,
  [arXiv:1103.4617 [hep-ph]], to appear in Phys. Rev. Lett. 
  
 \bibitem{Caswell:1985ui}
  W.~E.~Caswell and G.~P.~Lepage,
  Phys.\ Lett.\  B {\bf 167}, 437 (1986).

\bibitem{Kinoshita:1995mt}
  T.~Kinoshita and M.~Nio,
  Phys.\ Rev.\  D {\bf 53}, 4909 (1996).

\bibitem{Manohar:1997qy}
  A.~V.~Manohar,
  Phys.\ Rev.\  D {\bf 56}, 230 (1997)
    [arXiv:hep-ph/9701294].
  
  \bibitem{Carlson:2011zd}
  C.~E.~Carlson and M.~Vanderhaeghen,
    Phys.\ Rev.\  A {\bf 84}, 020102 (2011)
  [arXiv:1101.5965 [hep-ph]].

\end{thebibliography}
\end{document}